\documentclass[12pt]{article}
\pdfoutput=1
\usepackage{color}
\definecolor{darkblue}{rgb}{0.1,0.1,.7}
\usepackage[colorlinks, linkcolor=darkblue, citecolor=darkblue, urlcolor=darkblue, linktocpage]{hyperref}
\usepackage[]{amsmath}
\usepackage[]{graphicx}
\usepackage[]{latexsym}
\usepackage{slashed,graphicx,color,amsmath,amssymb}
\usepackage[mathscr]{eucal}
\usepackage{mathrsfs}
\usepackage{geometry}
\usepackage[margin=10pt,font=small,labelfont=bf]{caption}
\usepackage{amscd}
\usepackage{bm}
\usepackage{xcolor}
\usepackage{upgreek}
\usepackage[square, comma, sort&compress,numbers]{natbib}
\usepackage[all,cmtip]{xy}
\numberwithin{equation}{section}
\newcommand{\tr}{\mathrm{Tr}\,}

\newcommand{\rD}{\mathrm{D}}

\newcommand{\rI}{\mathrm{I}}

\begin{document}
\vspace*{-.6in} \thispagestyle{empty}
\vspace{.2in} {\Large
\begin{center}
{\bf Entanglement Entropy in Integrable Field Theories\\ with Line Defects I. Topological Defect}
\end{center}}
\vspace{.2in}
\begin{center}
Yunfeng Jiang
\\
\vspace{.3in}
\small{\textit{Institut f{\"u}r Theoretische Physik,
ETH Z{\"u}rich}},\\
\small{\textit{
Wolfgang Pauli Strasse 27,
CH-8093 Z{\"u}rich, Switzerland}
}

\end{center}

\vspace{.3in}

\begin{abstract}
\normalsize{In this paper and a companion one \cite{Jiang:defec2}, we study the effect of integrable line defects on entanglement entropy in massive integrable field theories in 1+1 dimensions. The current paper focuses on topological defects that are purely transmissive. Using the form factor bootstrap method, we show that topological defects do not affect the the entanglement entropy in the UV limit and modify slightly the leading exponential correction in the IR. This conclusion holds for both unitary and non-unitary field theories. In contrast, non-topological defects affect the entanglement entropy more significantly both in UV and IR limit and will be studied in the companion paper.}
\end{abstract}

\vskip 1cm \hspace{0.7cm}

\newpage

\setcounter{page}{1}
\begingroup
\hypersetup{linkcolor=black}
\tableofcontents
\endgroup

\section{Introduction}
\label{sec:1}
Quantum entanglement in many body systems is one of the most fundamental and fascinating phenomena in nature. It is therefore very important to \emph{quantify} (define proper quantities such as entanglement entropy, R\'enyi entropy, mutual information ect. to quantify quantum entanglement), \emph{calculate} (develop analytical and numerical methods to compute the quantities defined in the previous step for different systems) and \emph{measure} (in experiments) quantum entanglement. One most commonly studied quantity for quantum entanglement is the entanglement entropy. Computing entanglement entropy in many body systems is important as well as challenging for theories and also interesting for experiments (see for example \cite{Abanin,islam2015measuring}). In the past decade, impressive progress has been made in computing entanglement entropy, especially in conformal field theories where powerful geometric methods \cite{Holzhey:1994we,Calabrese:2004eu} and holographic insights \cite{Ryu:2006bv,Ryu:2006ef} play a crucial role. On the other hand, computation of entanglement entropy in generic massive quantum field theories remains open. Among them integrable field theories in 1+1 dimensions is special. Due to integrability, they are more tractable analytically than generic field theories. In a series of works \cite{Cardy:2007mb,CastroAlvaredo:2008kh,CastroAlvaredo:2008pf,CastroAlvaredo:2009ub,Bianchini:2015uea}, the authors studied entanglement entropy in the context of integrable field theories using the form factor bootstrap approach. The method is based on a generalized version of the traditional form factor bootstrap program \cite{Smirnov:1992vz,Karowski:1978vz} in the context of replica theory.\par

The form factor method is non-perturbative and is especially powerful in the IR limit where the result can be approximated to high precision by the first few terms in the spectral expansion. In this way, the authors obtained the leading corrections in the IR for the entanglement entropy, which takes a remarkably universal form that only depends on the spectrum of the theory \cite{Cardy:2007mb}. On the other hand, higher order corrections depend on more details of the theory.\par

Given an integrable field theory, one can introduce additional structures such as boundaries and defects. In general, these additional structures will spoil the integrability of the theory. However, there exist special kinds of boundaries and defects that preserve integrability \cite{Ghoshal:1993tm,Delfino:1994nx,Delfino:1994nr,Bajnok:2009hp} which allow one to apply powerful integrability techniques. In \cite{CastroAlvaredo:2008pf}, the authors investigated Ising model with integrable boundary conditions. The main goal of the current work is to study the case of integrable line defects in integrable field theories.\par

There are at least two reasons to consider line defects in quantum field theories. Firstly, homogeneous systems are usually mathematical idealization and it is important to estimate the effect of defects on entanglement entropy. This question has been investigated extensively for spin models and conformal field theories, see for example \cite{Affleck2009,Saleur:2013pva,Couvreur2016,Guo:2015uwa,Sakai:2008tt,Brehm:2015lja,2012JPhA...45o5301P} and references therein. Secondly, defects often contain useful information about the structure of the bulk theory and thus classification and the study of the defects are of considerable theoretical interest \cite{Petkova:2000ip,Frohlich:2006ch,Quella:2006de,Kormos:2009sk,Gaiotto:2012np,Bachas:2001vj,Bowcock:2003dr,Bowcock:2005vs,Bajnok:2009hp,Bajnok:2013waa}.\par

The requirement of integrability imposes very stringent constraints on the possible defects. The situation for free theories with $S=\pm1$ and for interacting theories, namely those with non-trivial $S$-matrices, are quite different.

For interacting integrable field theories, the only allowed defects are either purely reflective (i.e. integrable boundaries) or purely transmissive (i.e. topological defects) \cite{Delfino:1994nr,CastroAlvaredo:2002fc}. For the latter case, the defect is topological in the sense that the stress tensor is continuous across the defect and one can move the defect freely without costing any energy as long as it does not cross any other operators. This kind of defect is simply characterized by a transmission amplitude $T_{ij}(\theta)$. Form factors of a local operator in interacting theories can usually be worked out in practice for states with few particles and closed formulae for states with any number of particles are not known in general. Therefore, for interacting theories, we restrict ourselves to the leading corrections to the entanglement entropy \cite{Cardy:2007mb} in the IR and show how the defect affects the universal correction of the bulk entanglement entropy for both unitary and non-unitary theories.\par

For free theories, integrable defects can be transmissive and reflective simultaneously. This kind of defect is characterized by a transmission amplitude $T_{ij}(\theta)$ and a reflection amplitude $R_{ij}(\theta)$ and is non-topological. On the one hand, there exist closed formula for the form factors of local operators (e.g. branch-point twist fields) for any number of particles \cite{Cardy:2007mb,Bianchini:2016mra,Blondeau-Fournier:2016rtu} due to Wick's theorem. Given a closed formula for form factors of branch-point twist fields, we can expect more than just finding leading corrections in the IR. One can also consider the UV limit and make contact with the underlying CFT. In this way, one can obtain the boundary or defect entropy \cite{Affleck:1991tk} which characterize the number of degrees of freedoms on the boundary or defect.\par

The effects of topological and non-topological defects on entanglement entropy are also quite different. Intuitively, topological defects are very ``soft'' and one does not expect them to modify the bulk entanglement entropy drastically. This is indeed the case. In the UV limit where the main contribution comes from particles with high energies, the defect becomes transparent and the UV behavior is not modified. In the IR limit, the effect of the defect enters the leading exponential correction via transmission amplitude, which also takes a universal form
\begin{align}
&\text{Bulk EE: }&  &S_A=-\frac{c}{3}\log(\varepsilon m)+U-\frac{1}{16}\int_{-\infty}^\infty e^{-2mr\cosh\theta}d\theta&\\\nonumber
&\text{Topological defect:}& &S_A^{\text{defect}}=-\frac{c}{3}\log(\varepsilon m)+U-\frac{1}{16}\int_{-\infty}^\infty \hat{T}(\theta,\alpha)^2\, e^{-2mr\cosh\theta}d\theta.
\end{align}
where $\hat{T}(\theta,\alpha)=T(\frac{i\pi}{2}-\theta,\alpha)$ is the Wick rotated transmission amplitude. Here for simplicity, we give the result for theories whose spectrum consist a single particle. The result can be generalized straightforwardly to theories with more kind of particles. The leading correction takes a universal form which depends on both the spectrum of the theory and the transmission amplitude that characterize the topological defect. The transmission amplitude depends on some parameter which we denote $\alpha$. This parameter can be related to the coupling constant in the Lagrangian formulation. We find that for certain range of the parameter $\alpha$, the contribution from topological defect is finite while for the parameters outside this range, the effect of defect can be completely neglected also in the IR limit.\par

For non-unitary theories like the scaling Lee-Yang model, the calculation of entanglement entropy is much more subtle due to non-unitarity \cite{Bianchini:2014uta,Bianchini:2015uea}. Nevertheless, we have the same conclusion that the UV limit is unmodified while the IR corrections are modified for some range of the parameter that characterizes the defect. The explicit expression is slightly more involved and is given in (\ref{eq:Sdefect_non-unitary}).\par

For non-topological defects in the free theories, on the other hand, the entanglement entropy is modified more significantly both in the UV and IR. The analysis for these defects are more technical and the results will be given in the companion paper \cite{Jiang:defec2}.\par

The rest of the paper is structured as the follows. In section \ref{sec:ldefect} we introduce integrable line defects in integrable field theories in 1+1 dimensions. In section \ref{sec:EEdefect} we review the method of form factor bootstrap in the context of entanglement entropy and generalize it to the defect case. In section \ref{sec:unitary} and \ref{sec:nonunitary}, we consider the leading corrections to the entanglement entropy in unitary and non-unitary field theories using Sinh-Gordon theory and Scaling Lee-Yang theory as main examples, respectively. Finally we conclude in section \ref{sec:conclude}.

\section{Line defects in integrable field theories}
\label{sec:ldefect}
In this section, we review line defects in integrable field theories. Our discussion mainly follows the seminal papers \cite{Delfino:1994nx,Delfino:1994nr} and \cite{Bajnok:2007jg,Bajnok:2009hp}.

\subsection{Line defects}
We consider an infinitely long line defect in a 1+1 dimensional integrable quantum field theory (IQFT). In the Lagrangian description, the defect can be introduced by adding a term which has support only on the defect line
\begin{align}
S=S_B+g\int d^2 r\, \delta(x)\mathcal{L}_\rD(\phi_i,\partial_y\phi_i).
\end{align}
where $S_B$ is the bulk action and $\mathcal{L}_\rD$ is the defect lagrangian. We further require the interaction between bulk and defect be consistent with the existence of infinitely many conserved charges. The defect of this type preserves integrability and is referred to as an integrable defect. \par

In the bootstrap framework, the bulk physics is characterized by elastic scattering processes of bulk particles. In the presence of the defect, we also take into account scattering of bulk particles on the defect. Scattering processes on the defect is necessarily elastic due to integrability. A particle which scatters on the defect can either go through without changing its energy and momentum, or it can be reflected which preserves energy but flips the sign of momentum. It has been shown \cite{Delfino:1994nr,CastroAlvaredo:2002fc} that the only theories which allows both transmissive and reflective processes are the generalized free models (free boson and free fermion) with bulk scattering matrix $S=\pm 1$. For interacting integrable field theories, the defect has to be either purely reflective or purely transmissive. The first case is nothing but the integrable boundary, which has been investigated intensely starting from the seminal work of Ghoshal and Zamolodchikov \cite{Ghoshal:1993tm}. The latter case is called topological defect since the stress energy tensor is continuous across the defect. Topological defects have many special properties. They can be moved freely without changing physical observables and their positions are not important as long as they do not cross other local operators. One can also fuse a topological defect with integrable boundaries to obtain new integrable boundary conditions \cite{Bajnok:2007jg}. Topological defects of integrable field theories can be obtained as chiral perturbation of topological defects of the underlying conformal field theory \cite{Konik:1997gx,Bajnok:2013waa}.

\subsection{Defect algebra}
It is convenient to represent the defect formally as an operator $D$ \cite{Delfino:1994nr}. In an IQFT, excitations can be described by the so-called Zamolodchikov-Faddeev (ZF) operators $A^i(\theta)$ and $A_i^\dagger(\theta)$ which satisfy a set of algebraic relations (ZF algebra)
\begin{align}
A^i(\theta_1)A^j(\theta_2)=&\,S^{ij}_{kl}(\theta_1-\theta_2)A^l(\theta_2)A^k(\theta_1),\\\nonumber
A_i^\dagger(\theta_1)A_j^\dagger(\theta_2)=&\,S_{ij}^{kl}(\theta_1-\theta_2)A_l^\dagger(\theta_2)A_k^\dagger(\theta_1).
\end{align}
where $\theta_{1,2}$ are the rapidity of the particles and the index ``$i,j,k,l$'' denote quantum numbers of the excitations. Here $S_{ij}^{kl}(\theta_1-\theta_2)$ is the bulk scattering matrix which characterize the bulk dynamics. The consistency of ZF algebra leads to Yang-Baxter equations and unitarity conditions for the scattering matrix. Similarly, the interaction between the defect and bulk excitations can be described by an algebra between the defect operator $D$ and the ZF operators $A_i^\dagger(\theta)$. Assuming the defect has no internal degree of freedom, the defect algebra reads
\begin{align}
\label{eq:algebra}
A_i^\dagger(\theta)D=&\,R_{ij}^-(-\theta)A_j^\dagger(-\theta)D+T_{ij}^-(\theta)D A_j^\dagger(\theta)\\\nonumber
D A_i^\dagger(\theta)=&\,R_{ij}^+(\theta)D A_j^\dagger(-\theta)+T_{ij}^+(-\theta)A_j^\dagger(\theta)D.
\end{align}
where $R_{ij}^\pm$, $T_{ij}^\pm$ are the reflection and transmission matrices respectively. Similar to the bulk case, the consistency of the defect algebra gives rise to constraints for $R_{ij}^\pm$ and $T_{ij}^\pm$ \cite{Delfino:1994nr,CastroAlvaredo:2002fc}.

\subsection{Correlation functions}
Using the replica trick, the computation of entanglement entropy can be recast in terms of correlation functions of branch-point twist fields in the replica theory. In the bulk case, correlation functions of local operators can be computed by the form factor approach \cite{Smirnov:1992vz,Karowski:1978vz}. The main idea is to write correlation functions in terms of form factors via spectral expansion. The form factors are determined by solving a set of functional equations called the bootstrap axioms.\par

In the presence of a line defect, if the local operator does not sit on the defect, the form factor of the operator are essentially the same as the bulk case \cite{Bajnok:2009hp}. On the other hand, the presence of the defect affects the spectral expansion of correlation functions and gives rise to extra factors which are matrix elements of the defect operator.\par

In order to see this, it is more convenient to perform a ``double Wick rotation'' which exchanges temporal and spatial directions \cite{Delfino:1994nr,CastroAlvaredo:2002dj}. Originally, the defect is placed at $x=0$ which is localized in space but extend in time. After the Wick rotation, we place the defect at $t'=t_\rD$. Then the defect can be taken into account by defining an extended operator $\mathcal{D}$\footnote{Note that $\mathcal{D}$ here is different from the formal operator $D$ in (\ref{eq:algebra}).} placed at $t'=t_\rD$ acting on the bulk states. This is similar to the Wick rotation in the boundary integrable theories which defines a boundary state $|B\rangle$ \cite{Ghoshal:1993tm}. The correlation functions can be expressed as
\begin{align}
\langle \mathcal{O}_1(x_1,t_1)\cdots \mathcal{O}_n(x_n,t_n)\rangle_\rD=\frac{\langle0|T\left[\mathcal{O}_1(x_1,t_1)\cdots\mathcal{D}\cdots\mathcal{O}_n(x_n,t_n)\right]|0\rangle}
{\langle0|\mathcal{D}|0\rangle}
\end{align}
Here we put a ``$\rD$'' on the l.h.s to remind that it is the correlation functions in a defect theory. The fields on the l.h.s. are the fields in Heisenberg representation whose evolution is governed by the complete Hamiltonian (bulk+boundary). On the r.h.s. the operators $\mathcal{O}_i(x_i,t_i)$ are the bulk fields whose time evolution is governed by the bulk Hamiltonian.\par

Similar to the boundary case, we can represent the operator $\mathcal{D}$ in terms of ZF operators \cite{CastroAlvaredo:2002dj,Bajnok:2004jd}
\begin{align}
\mathcal{D}=\exp\left(\frac{1}{2}\int_{-\infty}^\infty \frac{d\theta}{2\pi}\mathscr{D}(\theta)  \right)
\end{align}
with
\begin{align}
\mathscr{D}(\theta)=&\,\hat{R}^-_{ij}(\theta)A^\dagger_i(-\theta)A^\dagger_j(\theta)+\hat{R}_{ij}^+(\theta)^*A^i(\theta)A^j(-\theta)+\\\nonumber
&\,\hat{T}_{ij}^-(\theta)A_i^\dagger(\theta)A_j(\theta)+\hat{T}_{ij}^+(\theta)^*A_i^\dagger(-\theta)A^j(-\theta)
\end{align}
where summation over repeated indices is understood. Here $\hat{T}^\pm_{ij}(\theta)=T^\pm_{ij}(\frac{i\pi}{2}-\theta)$ and $\hat{R}^\pm_{ij}(\theta)=R^\pm_{ij}(\frac{i\pi}{2}-\theta)$. After the Wick rotation we can perform the spectral expansion on the r.h.s. and apply the form factor bootstrap approach. For our purpose, the two-point function is most relevant. We consider the case where the two operators resides on two sides of the defect with $x_1=x_2=0$ and $t_1<t_\rD<t_2$. Schematically,
\begin{align}
\label{eq:spectral}
G_\rD(t_1,t_2)=&\,\langle0|\mathcal{O}_1(t_1)\,\mathcal{D}\,\mathcal{O}_2(t_2)|0\rangle=\sum_{M,N}\langle0|\mathcal{O}_1(t_1)|M\rangle\langle M|\mathcal{D}|N\rangle
\langle N|\mathcal{O}_2(t_2)|0\rangle\\\nonumber
=&\,\sum_{M,N}F^{\mathcal{O}_1}_{M}(F^{\mathcal{O}^\dagger_2}_N)^* e^{-(E_M\,r_L+E_N\,r_R)} \rD_{M,N}.
\end{align}
where $r_L=|t_1-t_\rD|$ and $r_R=|t_2-t_\rD|$. $F^{\mathcal{O}_i}_N$ are the $N$-particle form factor of operator $\mathcal{O}_i$, which can be determined by the form factor bootstrap program. Note that here we use a shorthand notation for the resolution of identity
\begin{align}
\mathbb{I}=\sum_N|N\rangle\langle N|\equiv \sum_{N=0}^\infty\frac{1}{N!}\int\frac{d\theta_1}{2\pi}\cdots\frac{d\theta_N}{2\pi}
|\theta_1,\cdots,\theta_N\rangle\langle\theta_1,\cdots,\theta_N|
\end{align}
The effect of the defect comes in through the matrix element $\rD_{M,N}$.\par

If we take a trivial defect $\rD_{M,N}=\delta_{M,N}$ in (\ref{eq:spectral}), we obtain the usual form factor expansion in the bulk theory. If we assume the defect to be purely reflective, then the non-vanishing matrix elements are $\rD_{2M,2N}=\rD_{2M,0}\rD_{0,2N}$. Note that
\begin{align}
\mathcal{D}|0\rangle=\exp\left(\frac{1}{4\pi}\int_{-\infty}^{\infty}\hat{R}_{ij}^-(\theta)A^\dagger_i(-\theta)A^\dagger_j(\theta)\,d\theta  \right)|0\rangle\equiv|\mathcal{B}\rangle
\end{align}
is nothing but the Ghoshal-Zamolodchikov boundary state. The spectral expansion of the two-point function in this case reduces to the product of two boundary expansions.
\begin{align}
G_\rD^\text{R}(t_1,t_2)=&\,\left(\sum_{M}F^{\mathcal{O}_1}_{2M}\,\rD_{2M,0}\,e^{-E_{2N}r_L}\right)
\left(\sum_N\left(F^{\mathcal{O}^\dagger_2}_{2N}\right)^*\,\rD_{0,2N}\,e^{-E_{2N}r_R}\right)\\\nonumber
=&\,\langle0|\mathcal{O}(t_1)|\mathcal{B}\rangle\langle\mathcal{B}|\mathcal{O}(t_2)|0\rangle.
\end{align}
The matrix elements $\rD_{2N,0}$ and $\rD_{0,2N}$ only depend on the reflection amplitude $R_{ij}$. If the defect is purely transmissive, the non-vanishing elements are $\rD_{N,N}$ and hence the spectral expansion becomes
\begin{align}
G_\rD^{\text{T}}(t_1,t_2)=\sum_N F^{\mathcal{O}_1}_N\,\left(F_N^{\mathcal{O}^\dagger_2}\right)^*\,\rD_{N,N}\,e^{-E_Nr}
\end{align}
where $r=|t_1-t_2|=r_L+r_R$ and the matrix element $\rD_{N,N}$ only depends on the transmission amplitude $T_{ij}$.

\section{Entanglement entropy in the presence of a line defect}
\label{sec:EEdefect}
\begin{figure}[h!]
\begin{center}
\includegraphics[scale=0.5]{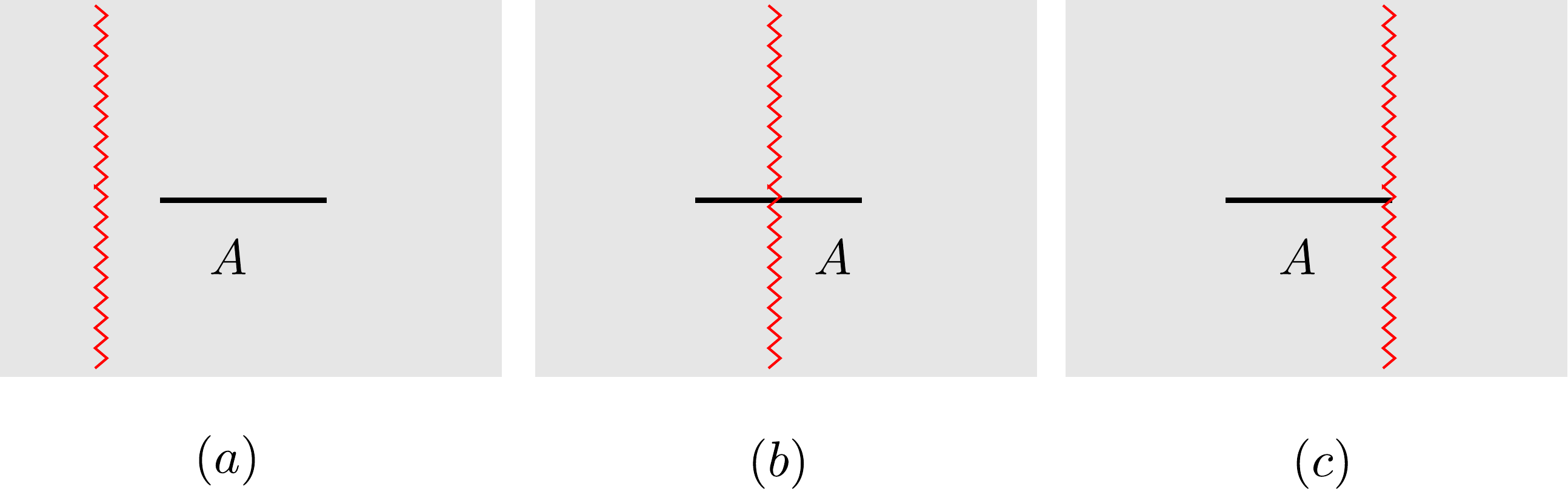}
\caption{The three possible positions of defects. The black line denote the interval $A$ and the red zigzag line stands for the defect. In this paper, we consider the case (b) where the defect sits in the interval $A$.}
\label{fig:defectPos}
\end{center}
\end{figure}

In this section, we formulate our set-up of computing the bipartite entanglement entropy in the presence of an integrable line defect. From now on, we focus on topological defects. We consider a line interval $A$ of length $r$ and denote its complement as $\bar{A}$. There are three possibilities for the location of the line defect, as is shown in figure\,\ref{fig:defectPos}.
In this paper, we consider the case (b) of figure\,\ref{fig:defectPos}. For case (a), the defect is outside the finite interval $A$ and can be moved to infinity since it is topological, which reduces to the bulk case. The case (c) is more subtle and will be considered elsewhere. Let us denote the Hilbert space by $\mathcal{H}=\mathcal{H}_A\otimes\mathcal{H}_{\bar{A}}$. For a given pure state $|\psi\rangle$, which in our case is the defect vacuum state $|0\rangle_\rD$, the reduced density matrix $\rho_A$ is given by
\begin{align}
\rho_A=\tr_{\mathcal{H}_{\bar{A}}}\left(|\psi\rangle\langle\psi|\right)
\end{align}
and the entanglement entropy is defined as the von Neumann entropy
\begin{align}
S_A=-\tr_{\mathcal{H}_A}\left(\rho_A\log\rho_A \right).
\end{align}

\subsection{Replica trick}
\begin{figure}[h!]
\begin{center}
\includegraphics[scale=0.4]{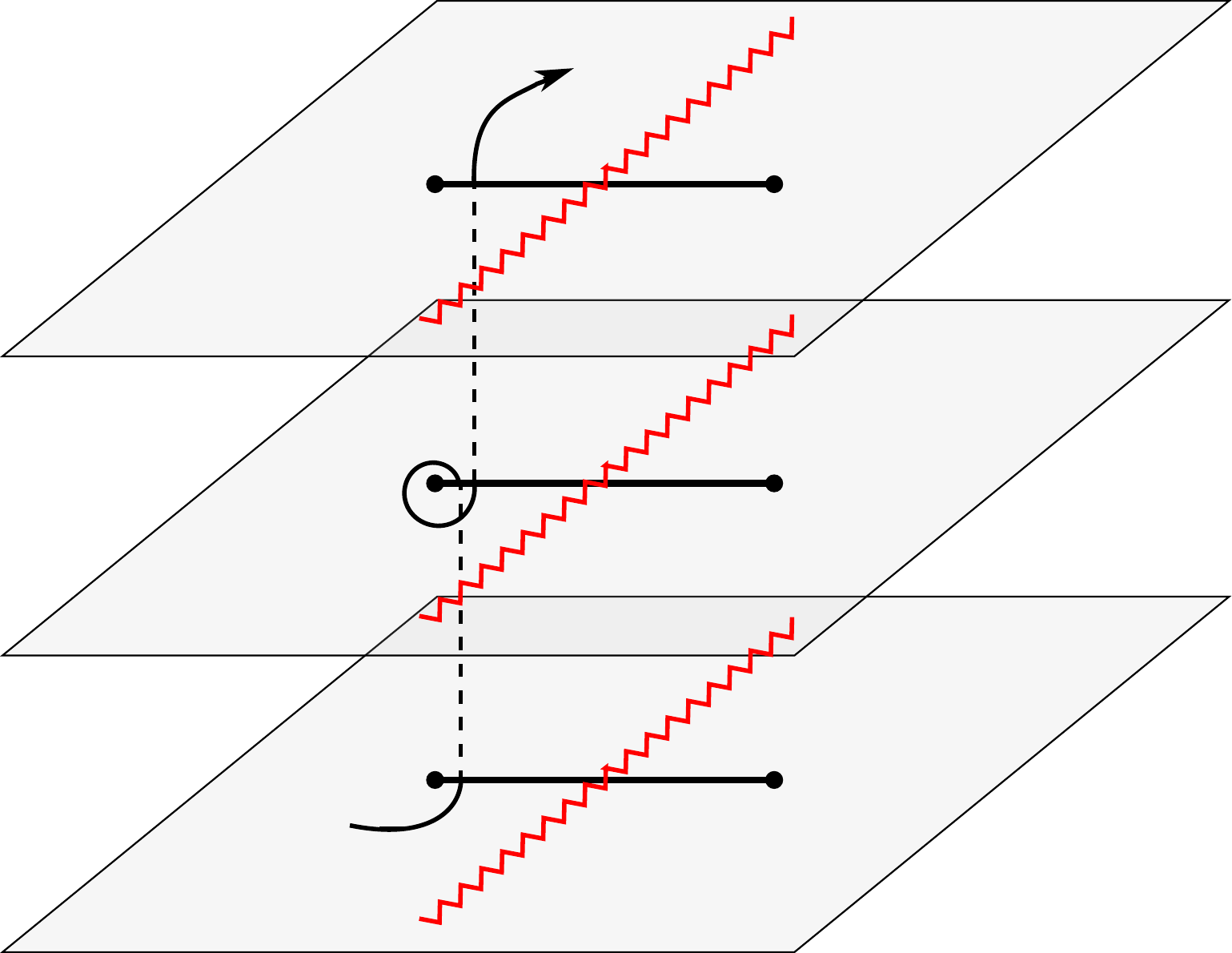}
\caption{Multiple copies of the original theory glued together. Here we take $n=3$. The dashed lines are the integrable line defects and the black lines are branch cuts.}
\label{fig:defect2}
\end{center}
\end{figure}

\begin{figure}[h!]
\begin{center}
\includegraphics[scale=0.4]{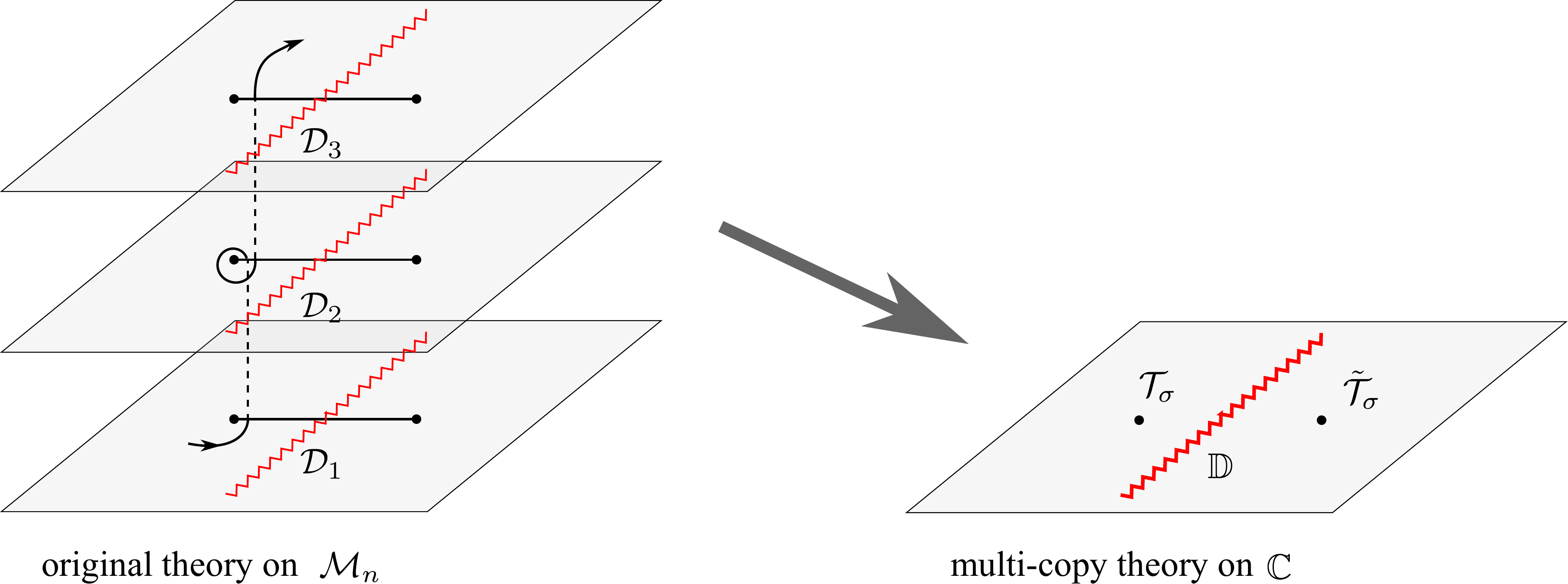}
\caption{From original theory on the Riemann surface $\mathcal{M}_n$ to the multi-copy theory on $\mathbb{C}$. The boundary conditions for the fields of the multi-copy theory can be taken into account by introducing branch-point twist fields.}
\label{fig:copy}
\end{center}
\end{figure}

We apply the replica trick to compute the entanglement entropy \cite{Cardy:2007mb}. One first compute $\tr\rho_A^n$ for positive integer $n$ and then analytically continue $n$ to any value. The entanglement entropy is obtained the as the following limit
\begin{align}
S_A=-\lim_{n\to 1}\frac{d}{d n}\tr\rho_A^n.
\end{align}
In order to compute $\tr\rho_A^n$, we consider $n$ copies of the original model and glue them cyclicly, as is shown in figure\,\ref{fig:defect2}.
This amounts to compute the partition function on the multi-sheeted Riemann surface with defects. The theory defined on this Riemann surface is however not local. Locality can be restored by introducing a multi-copy model formed by $n$ identical copies of the original model with specific boundary conditions for the fields. The problem is thus reduced to the computation of the partition function for the multi-copy model on the complex plane $\mathbb{C}$ with proper boundary conditions for the fields. When applying the replica trick, the defect is also duplicated and present on each sheet of the Riemann surface, which we denote by $\mathcal{D}_\mu$, $\mu=1,2,\cdots,n$. As a result, in the multi-copy model on the complex plane, the defect is the fused defect from all the $n$ copies, namely $\mathbb{D}=\mathcal{D}_1 \mathcal{D}_2 \cdots \mathcal{D}_n$. This is depicted in figure\,\ref{fig:copy}.
It has been shown that the entanglement entropy in the replica theory can be computed as the two-point function of certain special local operators called \emph{branch point twist fields} \cite{Cardy:2007mb}. Note that the prescription is different for non-unitary theories which will be given in section\,\ref{sec:nonunitary}. Using the branch point twist fields, we can write
\begin{align}
\label{eq:EET}
\tr_{\mathcal{H}_A}\rho_A^n=\mathcal{Z}_n\,\varepsilon^{4\Delta_n}\,\langle0|\mathcal{T}(t_1)\,\mathbb{D}\,\tilde{\mathcal{T}}(t_2)|0\rangle
\end{align}
Here $\mathcal{T}$ is the branch-point twist field and $\tilde{\mathcal{T}}=\mathcal{T}^\dagger$ is its hermitian conjugate. The constant $\mathcal{Z}_n$ is an $n$-dependent non-universal constant which satisfies $\mathcal{Z}_1=1$. The short-distance cut-off $\varepsilon$ is chosen in such a way that $\partial_n\mathcal{Z}_n|_{n=1}=0$. $\Delta_n$ is the conformal dimension of the branch point twist field
\begin{align}
\Delta_n=\frac{c}{24}\left(n-\frac{1}{n}\right).
\end{align}
The insertion of a topological defect does not modify the form of the two-point function in the UV limit, namely
\begin{align}
\label{eq:UV2pt}
\lim_{r_L,r_R\to0}\langle\mathcal{T}(t_1)\mathbb{D}\tilde{\mathcal{T}}(t_2)\rangle\sim\frac{1}{r^{4\Delta_n}}
\end{align}
where we recall that $r_L=|t_1-t_\rD|$, $r_R=|t_1-t_\rD|$ and $r=|t_1-t_2|$. This can be proven within the framework of form factor method, which will be present in section\,\ref{sec:short}. Plugging (\ref{eq:UV2pt}) into (\ref{eq:EET}) one finds that the r.h.s. is dimensionless and the entanglement entropy in the UV limit is given by
\begin{align}
S_A=-\frac{\partial}{\partial n}\left.\left(\mathcal{Z}_n (\varepsilon/r)^{4\Delta_n}\right)\right|_{n=1}=\frac{c}{3}\log\left(r/\varepsilon\right)
\end{align}
Therefore the entanglement entropy is not modified by the topological defect in the UV limit. Let us denote the mass of the lightest particle of the IQFT by $m$ and write
\begin{align}
\langle0|\mathcal{T}(t_1)\mathbb{D}\tilde{\mathcal{T}}(t_2)|0\rangle=\langle\mathcal{T}\rangle^2(1+X_\rD(n,r))
\end{align}
where we have chosen the normalization such that $\langle\mathbb{D}\rangle=1$. The entanglement entropy in the IR limit where $r_L,r_R\to\infty$ can then be written as
\begin{align}
S_A=-\lim_{n\to 1}\frac{d}{dn}\left(\mathcal{Z}_n(\varepsilon m)^{4\Delta_n}\right)\left(m^{-4\Delta_n}\langle\mathcal{T}\rangle^2\right)\left(1+X_{\rD}(n,r)\right)
\end{align}
In the limit $n\to 1$, the three parts are given by
\begin{align}
\left.\mathcal{Z}_n(\varepsilon m)^{4\Delta_n}\right|_{n=1}=
\left.m^{-4\Delta_n}\langle\mathcal{T}\rangle^2\right|_{n=1}=
\left.1+X_\rD(n,r)\right|_{n=1}=1.
\end{align}
The derivatives of the first two terms are
\begin{align}
\left.-\frac{d}{dn}\left(\mathcal{Z}_n(\varepsilon m)^{4\Delta_n}\right)\right|_{n=1}=-\frac{c}{3}\log(\varepsilon m)
\end{align}
and
\begin{align}
\left.-\frac{d}{dn}\left(m^{-4\Delta_n}\langle\mathcal{T}\rangle^2\right)\right|_{n=1}=U
\end{align}
where $U$ is a model dependent constant. Therefore the entanglement entropy takes the form in the IR limit
\begin{align}
\label{eq:EEIR}
S_A=-\frac{c}{3}\log(\varepsilon m)+U+\left.\partial_nX_{\rD}(n,r)\right|_{n=1}.
\end{align}
In the following, we will compute the leading corrections for different models. The effect of topological defect will enter via the last term of (\ref{eq:EEIR}).

\subsection{Spectral expansion and form factors of branch-point twist fields}
The main calculation of the entanglement entropy now boils down to the computation of the two-point function of the branch-point twist fields with defect operator $\langle0|\mathcal{T}(t_1)\mathbb{D}\tilde{\mathcal{T}}(t_2)|0\rangle$. Performing the spectral expansion we can write
\begin{align}
\label{eq:spec1}
\langle0|\mathcal{T}(t_1)\,\mathbb{D}\,\tilde{\mathcal{T}}(t_2)|0\rangle
=\sum_{M,N=0}^\infty f_{M,N}
\end{align}
where
\begin{align}
\label{eq:spec2}
f_{M,N}=&\,\frac{1}{M!N!}\sum_{j_1,\cdots,j_M=1}^n\sum_{k_1,\cdots,k_N=1}^n\int_{-\infty}^{\infty}
\prod_{r=1}^M\frac{d\theta_r}{2\pi}\prod_{s=1}^N\frac{d\theta'_s}{2\pi}\,\rD_{M,N}\\\nonumber
&\times {F}_M^{\mathcal{T}|j_1\cdots j_M}(\theta_1,\cdots,\theta_M)\, {F}^{\tilde{\mathcal{T}}|k_1\cdots k_N}_N(\theta'_1,\cdots,\theta'_N)^*
\times e^{-(mr_L\sum_{r=1}^M\cosh\theta_r+mr_R\sum_{s=1}^N\cosh\theta'_s)}.
\end{align}
Here the form factors ${F}_n^{\mathcal{T}}$ and ${F}_n^{\tilde{\mathcal{T}}}$ are the same as the bulk case and have been derived in \cite{Cardy:2007mb}. The factors $\rD_{M,N}$ are matrix elements of the defect operator
\begin{align}
\label{eq:DMM}
\rD_{M,N}=\langle\theta_1,\cdots,\theta_M|\mathbb{D}|\theta'_1,\cdots,\theta'_N\rangle.
\end{align}
where the asymptotic states are given by
\begin{align}
|\theta_1,\cdots,\theta_N\rangle_{\mu_1,\cdots,\mu_n}=A_{\mu_1}^\dagger(\theta_1)\cdots A_{\mu_N}^\dagger|0\rangle,\qquad \mu_k=1,\cdots,n
\end{align}
and for simplicity we omit the replica indices in (\ref{eq:DMM}) and in what follows.

For the interacting theories, the defect is purely transmissive. For simplicity, we consider the theory with one type of particle such as Sinh-Gordon and assume that the defect is parity invariant and does not have internal degrees of freedom. In this case, the bulk interaction is characterized by a simple $S$-matrix $S(\theta)$ and the defect is characterized by a transmission amplitude $T(\theta)$. The only non-vanishing matrix elements are
\begin{align}
\label{eq:dtrans}
\rD_{M,M}=&\,\langle\theta_1,\cdots,\theta_M|\mathbb{D}|\theta'_1,\cdots,\theta'_M\rangle\\\nonumber
=&\,\prod_{i=1}^M 2\pi \hat{T}(\theta_i)\delta_{\mu_i,\nu_i}\delta(\theta_i-\theta'_i)+\text{permutations}.
\end{align}
where $\mu_i,\nu_i=1,\cdots,n$ are replica indices of particles with rapidities $\theta_i,\theta'_i$ respectively. Here `permutations' means that we need to take into account all possible contractions between the incoming and outgoing particles. The contraction between particles $\{\theta_i,\mu_i\}$ and $\{\theta'_j,\nu_j\}$ contributes a factor $\hat{T}(\theta_i)\delta_{\mu_i,\nu_j}\delta(\theta_i-\theta'_j)$. When the order of the particles are exchanged, we pick up the corresponding $S$-matrix in the replica theory $[S(\theta)]^{\delta_{\mu_i,\mu_j}}$ since only the particles with the same replica index have non-trivial interactions. For example, the first two defect matrix elements are given by
\begin{align}
\label{eq:D22}
\rD_{1,1}=&\,\langle\theta_1|\mathbb{D}|\theta'_1\rangle=2\pi\hat{T}(\theta_1)\delta_{\mu_1,\nu_1}\delta(\theta_1-\theta'_1),\\\nonumber
\rD_{2,2}=&\,\langle\theta_1,\theta_2|\mathbb{D}|\theta'_1,\theta'_2\rangle\\\nonumber
=&\,4\pi^2\hat{T}(\theta_1)\hat{T}(\theta_2)\delta_{\mu_1,\nu_1}\delta_{\mu_2,\nu_2}\delta(\theta_1-\theta'_1)\delta(\theta_2-\theta'_2)\\\nonumber
+&\,4\pi^2\hat{T}(\theta_1)\hat{T}(\theta_2)
[S(\theta_1,\theta_2)]^{\delta_{\mu_1,\mu_2}}\delta_{\mu_2,\nu_1}\delta_{\mu_1,\nu_2}\delta(\theta_2-\theta'_1)\delta(\theta_1-\theta'_2)
\end{align}
Plugging the defect matrix element (\ref{eq:dtrans}) into the spectral expansion (\ref{eq:spec2}), we find that
\begin{align}
f_{M,M}=\frac{1}{M!}\sum_{j_1,\cdots,j_M=1}^n\int_{-\infty}^{\infty}
\prod_{r=1}^M\frac{d\theta_r}{2\pi}{\hat{T}(\theta_r)}
\times\left|{F}_M^{\mathcal{T}|j_1\cdots j_M}(\theta_1,\cdots,\theta_M)\right|^2
\times e^{-mr\sum_{r=1}^M\cosh\theta_r}.
\end{align}
As we see the effect of purely transmissive defect is simply modifying the measure of the integral by multiplication of transmission factors and only depends on $r$ instead of $r_L,r_R$. This confirms our claim that we can move topological defects freely within the interval. The same result can also be obtained from a slightly different point of view from the defect form factor axioms \cite{Bajnok:2009hp}.

\section{UV limit}
\label{sec:short}
In this section, we study the short distance behavior of the defect two-point function $\langle\mathcal{T}(t_1)\mathbb{D}\tilde{\mathcal{T}}(t_2)\rangle$ and show that it is the same as $\langle\mathcal{T}(t_1)\tilde{\mathcal{T}}(t_2)\rangle$ for topological defects, namely
\begin{align}
\lim_{r_L,r_R\to 0} \langle\mathcal{T}(t_1)\mathbb{D}\tilde{\mathcal{T}}(t_2)\rangle\approx\lim_{t_1\to t_2}\langle\mathcal{T}(t_1)\tilde{\mathcal{T}}(t_2)\rangle\sim\frac{1}{r^{4\Delta_n}}.
\end{align}
This implies that the UV limit of the entanglement entropy is not modified by topological defects.

Purely transmissive defects can be obtained from topological defects of the underlying CFT by chiral perturbation \cite{Konik:1997gx,Bajnok:2013waa}. It has been shown in \cite{Brehm:2015plf,Gutperle:2015kmw} that topological defects in CFT do not change the universal behavior of entanglement entropy. This can also be analyzed in the framework of form factor approach. On the other hand, if the defect is non-topological, then it will in general modify the UV behavior.\par

\begin{figure}[h!]
\begin{center}
\includegraphics[scale=0.5]{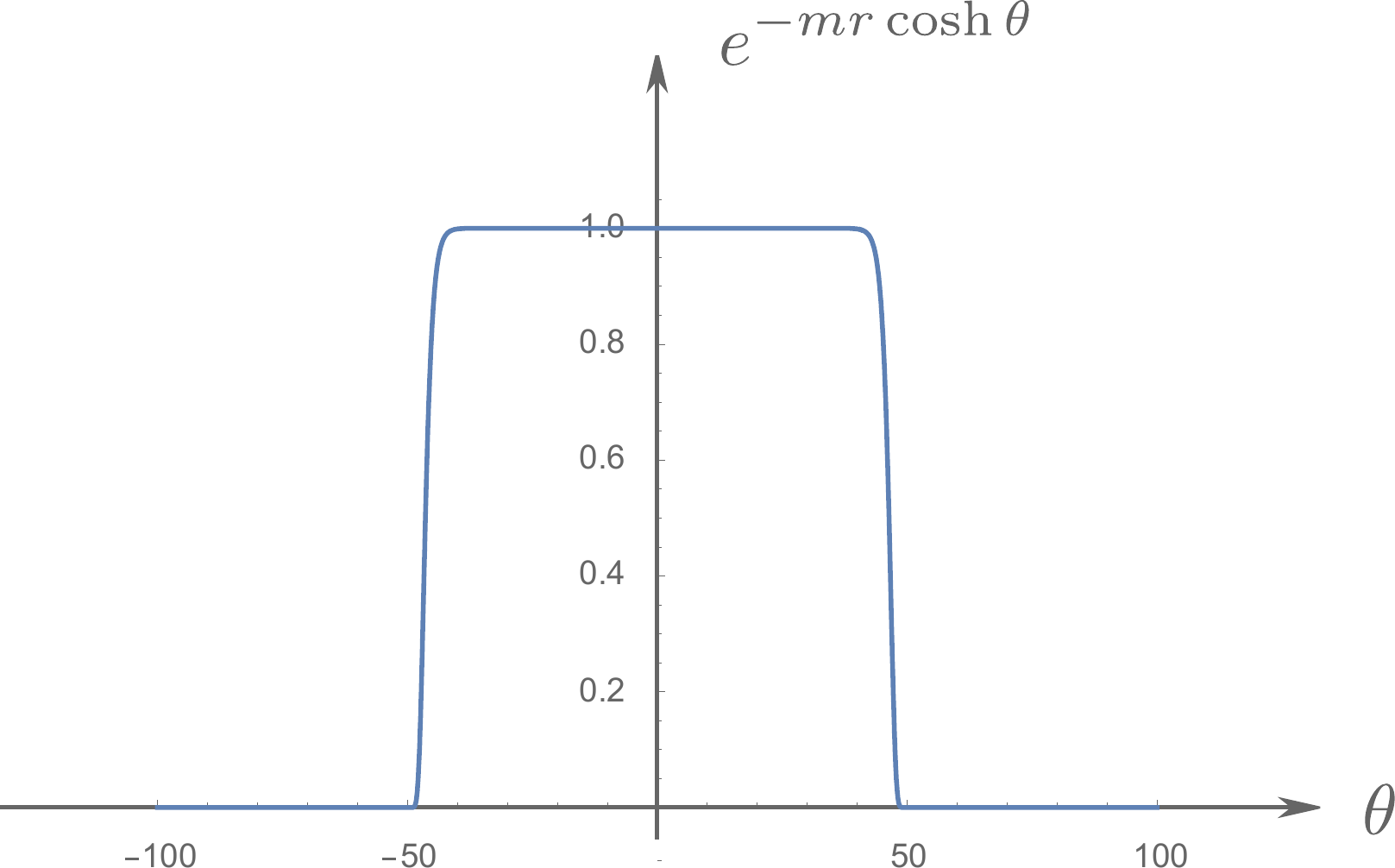}
\caption{The behavior of factor $e^{-mr\cosh\theta}$, here we take $mr=10^{-20}$, and $-\log(mr)\approx 46.05$.}
\label{fig:expcosh}
\end{center}
\end{figure}

In order to analyze the short distance behavior of two-point functions in the form factor framework, it is useful to consider the form factor expansion of the logarithm of the two-point function \cite{Smirnov:1990vm,Babujian:2003sc,Bianchini:2015uea}
\begin{align}
\log\left(\frac{\langle\mathcal{O}(r)\tilde{\mathcal{O}}(0)\rangle}{\langle\mathcal{O}\rangle^2}\right)
=&\,\sum_{k=1}^\infty\frac{1}{k!}\sum_{\mu_1,\cdots,\mu_k=1}^n\left(\prod_{j=1}^k\int_{-\infty}^{\infty}\frac{d\theta_j}{2\pi}\right)\\\nonumber
&\times H_k^{\mathcal{O}|\mu_1\cdots\mu_k}(\theta_1,\cdots,\theta_k)e^{-rm\sum_{j=1}^k\cosh\theta_j}
\end{align}
Similarly, the two-point function with a purely transmissive defect has the following expansion
\begin{align}
\label{eq:log2pt}
\log\left(\frac{\langle\mathcal{O}(r)\,\mathbb{D}\,\tilde{\mathcal{O}}(0)\rangle}{\langle\mathcal{O}\rangle^2}\right)=&\,
\sum_{k=1}^\infty\frac{1}{k!}\sum_{\mu_1,\cdots,\mu_k=1}^n\left(\prod_{j=1}^k\int_{-\infty}^{\infty}\frac{d\theta_j}{2\pi}\hat{T}(\theta_j)\right)\\\nonumber
&\times H_k^{\mathcal{O}|\mu_1\cdots\mu_k}(\theta_1,\cdots,\theta_k)e^{-rm\sum_{j=1}^k\cosh\theta_j}
\end{align}
where the function $H_k^{\mathcal{O}|\mu_1\cdots\mu_k}(\theta_1,\cdots,\theta_n)$ can be seen as the ``connected parts'' of the form factors and can be worked out explicitly for any $k$. The first few $H_k$'s are given by \cite{Bianchini:2015uea}\footnote{Note that here we assume the theory is unitary and the expansion is slightly different from that in \cite{Bianchini:2015uea} which is for the non-unitary case up to some signs.}
\begin{align}
H_1^{\mathcal{O}|\mu_1}(\theta)=&\,\langle\mathcal{O}_1\rangle^{-2}\,|F_1^{\mathcal{O}|\mu_1}(\theta)|^2,\\\nonumber
H_2^{\mathcal{O}|\mu_1\mu_2}(\theta_1,\theta_2)=&\,\langle\mathcal{O}\rangle^{-2}|F_2^{\mathcal{O}|\mu_1\mu_2}(\theta_1,\theta_2)|^2-
H_1^{\mathcal{O}|\mu_1}(\theta_1)H_1^{\mathcal{O}|\mu_2}(\theta_2)\\\nonumber
H_3^{\mathcal{O}|\mu_1\mu_2\mu_3}(\theta_1,\theta_2,\theta_3)=&\,\langle\mathcal{O}\rangle^{-2}|F_3^{\mathcal{O}|\mu_1\mu_2\mu_3}(\theta_1,\theta_2,\theta_3)|^2
-H_1^{\mathcal{O}|\mu_1}(\theta_1)H_1^{\mathcal{O}|\mu_2}(\theta_2)H_1^{\mathcal{O}|\mu_3}(\theta_3)\\\nonumber
&\,-H_2^{\mathcal{O}|\mu_1\mu_2}(\theta_1,\theta_2)H_1^{\mathcal{O}|\mu_3}(\theta_3)
-H_2^{\mathcal{O}|\mu_2\mu_3}(\theta_2,\theta_3)H_1^{\mathcal{O}|\mu_1}(\theta_1)\\\nonumber
&\,-H_2^{\mathcal{O}|\mu_1\mu_3}(\theta_1,\theta_3)H_1^{\mathcal{O}|\mu_2}(\theta_2).
\end{align}
By studying the limit $mr\to0$ one can obtain the scaling dimension $\Delta_{\mathcal{O}}$ of the operator $\mathcal{O}$, which is given by \cite{Smirnov:1990vm,Babujian:2003sc,Bianchini:2015uea}
\begin{align}
\Delta_{\mathcal{O}}=\sum_{k=1}^\infty\frac{1}{k!}\sum_{\mu_1,\cdots,\mu_k=1}^n\left(\prod_{j=1}^{k-1}\int_{-\infty}^{\infty}\frac{d\theta_j}{2\pi}\right)
H_k^{\mathcal{O}|\mu_1\cdots\mu_k}(\theta_1,\cdots,\theta_{k-1},0).
\end{align}
Now we analyze the same limit for the expansion with defect (\ref{eq:log2pt}). We shift the integration variable $\theta_j\to\theta_j+\theta_k$ ($j=1,\cdots,k-1$). Due to the relativistic invariance of the bulk form factor, we have
\begin{align}
\label{eq:spD}
&\frac{1}{k!}\sum_{\mu_1,\cdots,\mu_k=1}^n\left(\prod_{j=1}^k\int_{-\infty}^{\infty}\frac{d\theta_j}{2\pi}\hat{T}(\theta_j)\right)
H_k^{\mathcal{O}|\mu_1\cdots\mu_k}(\theta_1,\cdots,\theta_k)e^{-rm\sum_{j=1}^k\cosh\theta_j}\\\nonumber
&\,=\frac{1}{k!}\sum_{\mu_1,\cdots,\mu_k=1}^n\left(\prod_{j=1}^{k-1}\int_{-\infty}^{\infty}\frac{d\theta_j}{2\pi}\hat{T}(\theta_j+\theta_k)
\int_{-\infty}^{\infty}\hat{T}(\theta_k)\frac{d\theta_k}{2\pi}\right)
H_k^{\mathcal{O}|\mu_1\cdots\mu_k}(\theta_1,\cdots,\theta_{k-1},0)\\\nonumber
&\,\qquad \times e^{-rm\sum_{j=1}^{k-1}\cosh(\theta_j+\theta_k)+\cosh\theta_k}
\end{align}
Using the fact
\begin{align}
\sum_{j=1}^{k-1}\cosh(\theta_j+\theta_k)+\cosh\theta_k=&\,\cosh\theta_k\left(\sum_{j=1}^{k-1}\cosh\theta_j+1\right)+\sinh\theta_k
\left(\sum_{j=1}^{k-1}\sinh\theta_j  \right)\\\nonumber
=&\,\xi \cosh(\theta_k+\tau)
\end{align}
where
\begin{align}
\xi=\sqrt{\left(\sum_{j=1}^{k-1}\cosh\theta_j+1\right)^2-\left(\sum_{j=1}^{k-1}\sinh\theta_j  \right)^2}
\end{align}
and
\begin{align}
\cosh\tau=\frac{1}{\xi}\left(\sum_{j=1}^{k-1}\cosh\theta_j+1\right),\qquad \sinh\tau=\frac{1}{\xi}\left(\sum_{j=1}^{k-1}\sinh\theta_j  \right).
\end{align}
we can write the exponential factor in the second line of (\ref{eq:spD}) as $e^{-mr\cosh(\theta_k+\tau)}$. This factor is almost 1 in the interval $\log(mr)<\theta_k+\tau<-\log(mr)$ and zero outside the interval when $mr\to 0$, as is shown in figure\,\ref{fig:expcosh}.
Therefore the small $mr$ limit is determined by the large $\theta_k$ behavior of the integral. Since $\theta_k$ appears in every transmission amplitude $\hat{T}(\theta_j+\theta_k)$ and $\hat{T}(\theta_k)$, we thus need to analyze the large rapidity behavior for the transmission matrix $\hat{T}(\theta)$.\par

A purely transmissive defect is characterized by two transmission factors $T_\pm(\theta)$ (for parity invariant defect we have $T_+(\theta)=T_-(\theta)$) which describe the scattering of particles with defect from left and right, respectively. The two transmission factors are not independent and are related by defect unitarity and crossing symmetry
\begin{align}
\label{eq:Tpm}
T_+(-\theta)T_-(\theta)=1,\qquad T_-(\theta)=T_+(i\pi-\theta)
\end{align}
The quantity that appear in the spectral expansion is given by $\hat{T}(\theta)=T_-(\frac{i\pi}{2}-\theta)=T_+(\frac{i\pi}{2}+\theta)$. The first relation in (\ref{eq:Tpm}) can be written as $\hat{T}(\theta)\hat{T}(\theta-i\pi)=1$. When $\theta$ is very large, the shift of $i\pi$ can be neglected and we can write
\begin{align}
\lim_{\theta\to\pm\infty}\hat{T}(\theta)^2=1
\end{align}
Therefore asymptotically $\hat{T}(\theta)\to\pm 1$ by unitarity. Now let us specify to the models we study in this paper. For scaling Lee-Yang model, the transmission amplitude is given in (\ref{eq:TsLY}) with the asymptotic behavior $\lim_{\theta\to\pm\infty}\hat{T}_{\text{sLY}}(\theta)=1$. Therefore the defect is transparent in the large rapidity limit and does not modify the scaling dimension. For Sinh-Gordon model, the transmission amplitude is given in (\ref{eq:TShG}) with the asymptotic behavior $\lim_{\theta\to\pm\infty}\hat{T}_{\text{ShG}}(\theta)=\mp 1$. This might create some problem for the terms involve odd number of particles since the asymptotics of two limits cancel each other. However, Sinh-Gordon theory is a parity invariant theory and the operators can be classified according to the parity. For parity even (odd) operators, only terms with even (odd) number of particles contribute to the spectral expansion \cite{Fring:1992pt}. This property is still true in the replica theory and the branch-point twist operator is parity even, so the terms with odd number of particles are automatically zero \cite{CastroAlvaredo:2011zs}. For the terms with even number of particles, the products of transmission amplitudes are transparent in the large rapidity limit. To conclude, topological defects do not change the two-point function of branch-point twist fields in the UV and the entanglement entropy stays the same.

\section{Unitary case : Sinh-Gordon theory}
\label{sec:unitary}
\begin{figure}[h!]
\begin{center}
\includegraphics[scale=0.35]{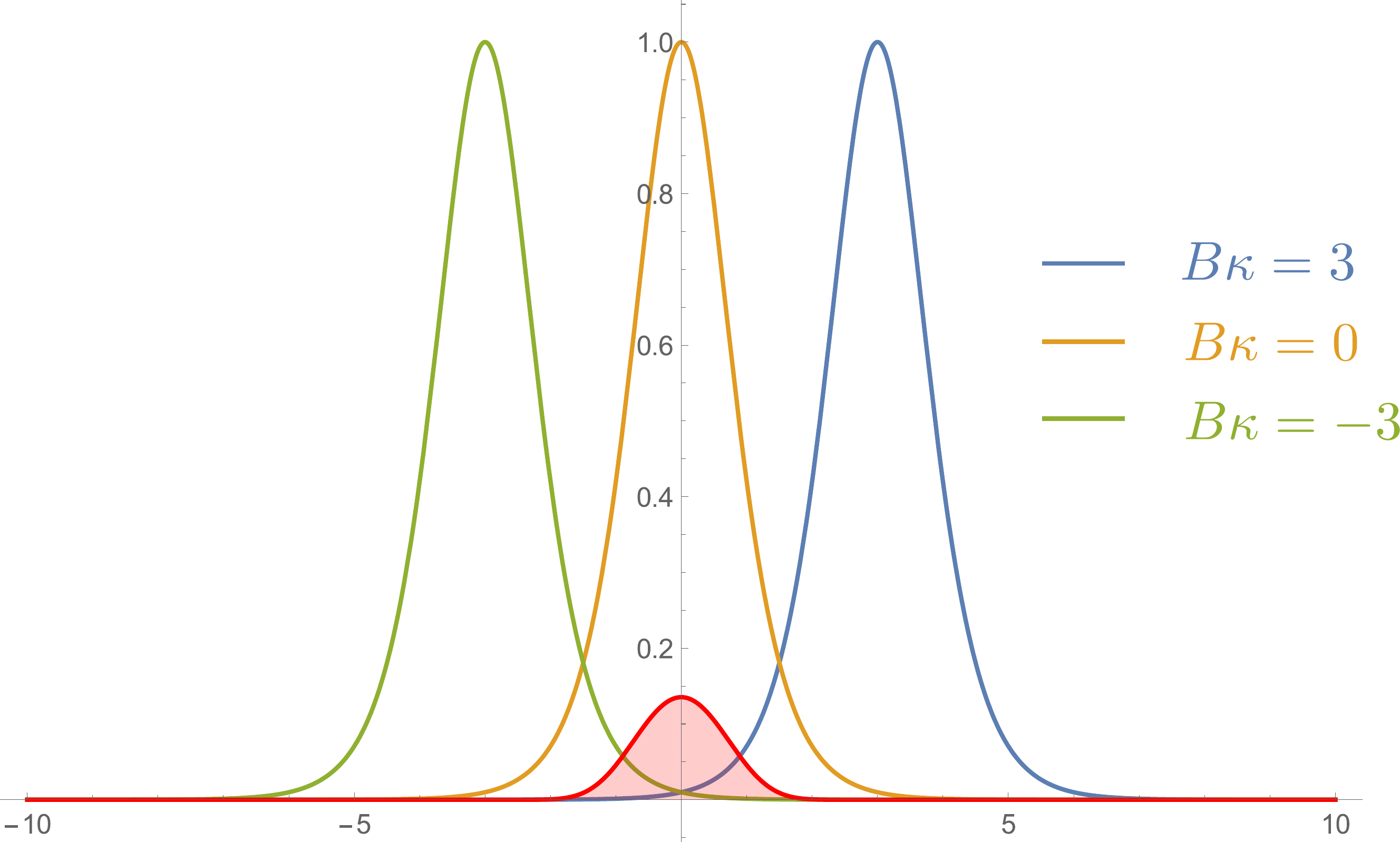}
\caption{Plot of $f_{\text{exp}}(\theta)$ and $f_\rD(\theta,B\kappa)$ with $B\kappa=0,\pm3$. The red shaded line denote $f_{\text{exp}}(\theta)$ with $mr=1$.}
\label{fig:overlap}
\end{center}
\end{figure}

In this section, we consider the leading correction to the entanglement entropy for unitary IQFT. We take the Sinh-Gordon theory as our example. There is only one type of particle and the scattering process is characterized by the following bulk scattering matrix\footnote{Note that the convention of the scattering matrix \cite{Cardy:2007mb} is slightly different from those in \cite{Bajnok:2007jg}. We follow the convention of \cite{Cardy:2007mb} in this paper.}
\begin{align}
S(\theta)=\frac{\tanh\frac{1}{2}\left(\theta-\frac{\pi}{2}B\right)}{\tanh\frac{1}{2}\left(\theta+\frac{\pi}{2}B\right)},\qquad B=\frac{2\beta^2}{\beta^2+1}
\end{align}
where $\beta$ is the coupling constant in the Sinh-Gordon lagrangian. The transmission amplitude is given by (see for example \cite{Bajnok:2007jg})
\begin{align}
\label{eq:TShG}
T(\theta)=-i\frac{\sinh\frac{1}{2}\left(\theta-\frac{i\pi}{2}+B\,\kappa\right)}{\sinh\frac{1}{2}\left(\theta+\frac{i\pi}{2}+B\,\kappa\right)},\qquad
\hat{T}(\theta)=T(\frac{i\pi}{2}-\theta)=\tanh\frac{1}{2}(\theta-B\,\kappa)
\end{align}
where $\kappa$ is a parameter that characterize the defect. In the Sinh-Gordon model, the leading contribution starts with $\rD_{2,2}$ since the one-particle form factor is vanishing. Plugging (\ref{eq:D22}) into the spectral expansion (\ref{eq:spec1}) and (\ref{eq:spec2}), we obtain the leading contribution
\begin{align}
f_{2,2}=&\,\frac{1}{2}\sum_{\mu_1,\mu_2=1}^n\int_{-\infty}^\infty \frac{d\theta_1 d\theta_2}{(2\pi)^2}\,\hat{T}(\theta_1)\hat{T}(\theta_2)
\left|F_2^{\mathcal{T}|\mu_1\mu_2}(\theta_1,\theta_2)\right|^2\,e^{-mr(\cosh\theta_1+\cosh\theta_2)}\\\nonumber
=&\,\frac{n}{2}\sum_{\mu=1}^n\int_{-\infty}^\infty \frac{d\theta_1 d\theta_2}{(2\pi)^2}\,\hat{T}(\theta_1)\hat{T}(\theta_2)
\left|F_2^{\mathcal{T}|1\mu}(\theta_1,\theta_2)\right|^2\,e^{-mr(\cosh\theta_1+\cosh\theta_2)}
\end{align}
where in the second line we have used the property of the two-particle form factor $F^{\mathcal{T}|i\,i+k}(\theta_1,\theta_2)=F^{\mathcal{T}|j\,j+k}(\theta_1,\theta_2)$ to perform the summation over $\mu_2$. The explicit form of the two-particle form factors of branch-point twist fields can be found in \cite{Cardy:2007mb}. For the leading correction, we do not need the explicit form. It is sufficient to apply the following important property of the two-particle form factor of the branch-point twist field after we analytically continue $n$ according to the prescription given in \cite{Cardy:2007mb}
\begin{align}
\label{eq:anacon}
\frac{\partial}{\partial n}\left.\sum_{\mu=1}^n\left|F_2^{\mathcal{T}|1\mu}(\theta_1,\theta_2)\right|^2\right|_{n=1}=\langle\mathcal{T}\rangle^2\frac{\pi^2}{2}\delta(\theta_1-\theta_2)
\end{align}
Using (\ref{eq:anacon}), the leading correction to entanglement entropy reads
\begin{align}
\label{eq:leading}
s_{2,2}=-\frac{\partial}{\partial n}\left[f_{2,2}\right]_{n=1}=-\frac{1}{16}\int_{-\infty}^{\infty}\hat{T}(\theta)^2\,e^{-2mr\cosh\theta}\,d\theta.
\end{align}
If we set $\hat{T}(\theta)\to 1$, we recover the leading correction to the bulk entanglement entropy. Note that our result for the leading exponential correction (\ref{eq:leading}) is also universal in the sense that it only depends on the spectrum of the theory and the transmission amplitude of the defect. The contribution due to the presence of the defect can be extract to be
\begin{align}
\delta s_\rD=-\frac{1}{16}\int_{-\infty}^{\infty}(\hat{T}(\theta)^2-1)\,e^{-2mr\cosh\theta}\,d\theta.
\end{align}
For Sinh-Gordon model, we plug in the explicit transmission amplitude and obtain
\begin{align}
\label{eq:deltasD}
\delta s_\rD=&\,-\frac{1}{16}\int_{-\infty}^{\infty}\left[\left(\tanh\frac{\theta-B\kappa}{2}\right)^2-1\right]\,e^{-2mr\cosh\theta}\,d\theta\\\nonumber
=&\,\frac{1}{16}\int_{-\infty}^{\infty}\frac{e^{-2mr\cosh\theta}}{\left(\cosh\frac{1}{2}(\theta-B\kappa)\right)^2}\,d\theta
\end{align}
This quantity depends on the parameter $B\kappa$. We plot the function $f_{\text{exp}}(\theta)=e^{-2mr\cosh\theta}$ and $f_\rD(\theta,B\kappa)=1/(\cosh\frac{1}{2}(\theta-B\kappa))^2$ with different values of $B\kappa$ in figure\,\ref{fig:overlap}.
We see that both functions have similar behavior, namely they are finite in some range and quickly damped away outside this range. The parameter $B\kappa$ controls the center of the ``peak''. If the finite range of $f_{\text{exp}}(\theta)$ and $f_\rD(\theta,B\kappa)$ do not overlap, the product is damped and gives almost zero. On the other hand, if the finite range of the two functions overlap, then the contribution is small but finite.\par

\begin{figure}[h!]
\begin{center}
\includegraphics[scale=0.25]{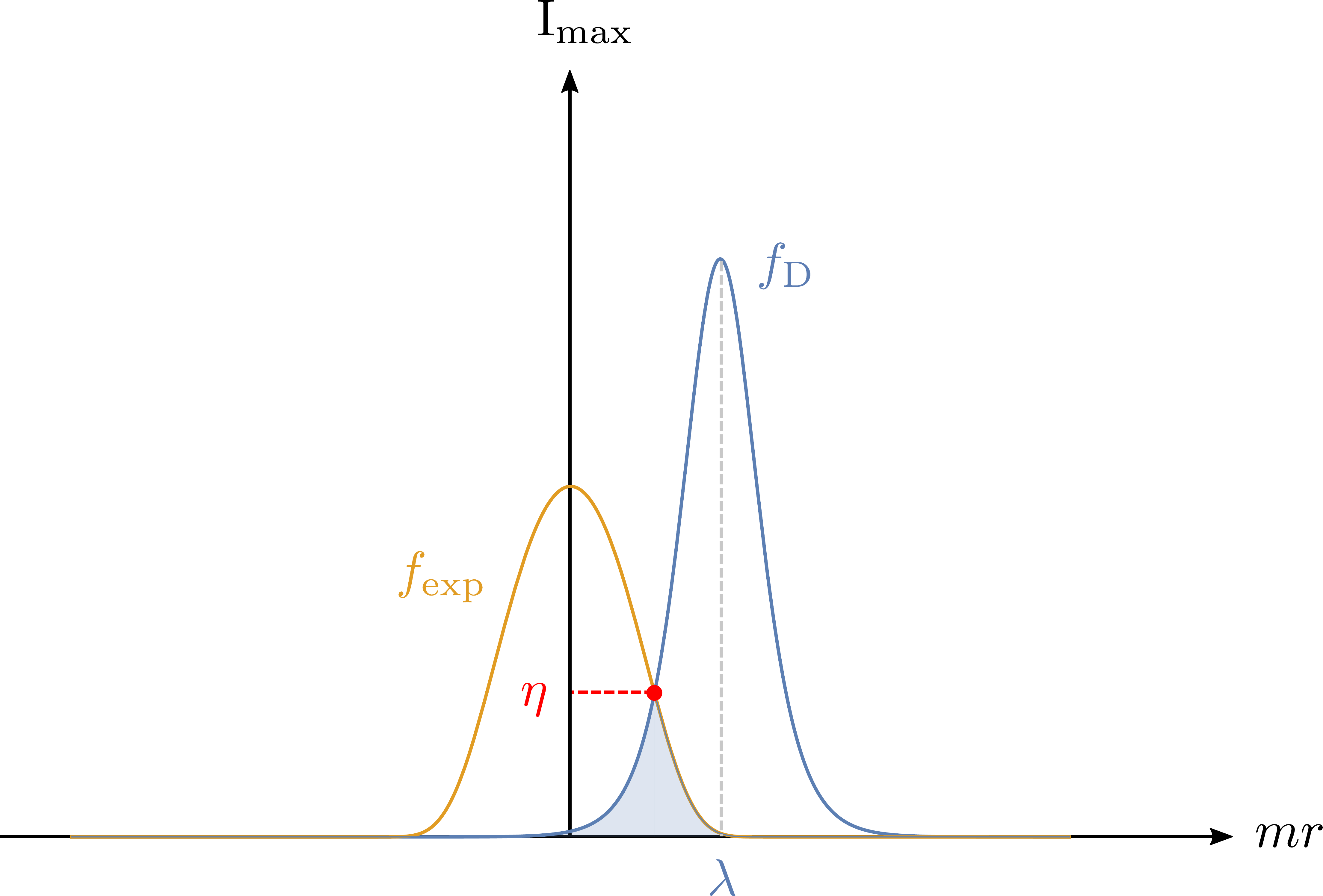}
\caption{Overlap of $f_{\text{exp}}$ and $f_{\text{D}}$. The unit of the two axis are $mr$ and $\text{I}_{\text{max}}$, respectively.}
\label{fig:peakoverlap}
\end{center}
\end{figure}

In order to have some idea about the finite range, we perform some numeric analysis in what follows. Let us consider the overlap of two functions $f_{\text{exp}}(\theta)$ and $f_{\text{D}}(\theta)$, which is the colored region in figure\,\ref{fig:peakoverlap}. To be more explicit, we consider
\begin{align}
f_{\text{exp}}(\theta)=e^{-2mr\cosh\theta},\qquad f_{\rD}(\theta)=\frac{1}{\cosh(\frac{1}{2}(\theta-\lambda\,mr))^2}
\end{align}
where we have set $B\kappa=\lambda\,mr$, namely we use $mr$ as the unit for the rapidity $\theta$. The value of $\lambda$ determines the position of the peak, as is shown in figure\,\ref{fig:peakoverlap}.
We denote the maximal value of $f_{\text{exp}}(\theta)$ as $\text{I}_{\text{max}}=e^{-2mr}$ and set it as the unit for the value of the two functions. The maximal overlap is obtained at $\lambda=0$. As $|\lambda|$ increases, the overlap decreases. For large enough $|\lambda|$, the curves of two functions intersect at one point\footnote{Both functions go to zero at large enough rapidity and as a result there are infinitely many points that are close to each other. These are not the intersecting point we mean here. There is a unique intersecting point at which both functions are finite and coincide.} which we denote by $\theta=\theta^*$. We denote the value of the two functions at this point by $f_{\text{exp}}(\theta^*)=f_{\text{D}}(\theta^*)=\eta^*\,\text{I}_{\text{max}}$. Both $\theta^*$ and $\eta^*$ depend on $\lambda$ and hence $\theta^*=\theta^*(\lambda)$ and $\eta^*=\eta^*(\lambda)$. Let us fix a small value for $\eta^*$, say $\eta_0=1.0\times 10^{-5}$ and we consider the overlap to be negligible for $0<\eta^*<\eta_0$ and non-negligible otherwise. The requirement for non-negligible overlap $1>\eta^*>\eta_0$ fixes a finite range for $|\lambda|<\lambda_0$ where $\lambda_0>0$ and satisfies $\eta^*(\lambda_0)=\eta_0^*$. The analytic relation $\eta^*(\lambda)$ is the solution of the following equation
\begin{align}
\eta^*e^{-2mr}=\frac{1}{\cosh\left(\frac{1}{2}\left[\text{arccosh}(1-\log\eta^*/2mr)-\lambda mr\right]  \right)^2}
\end{align}
and is hard to write down explicitly. However, for a fixed $\eta_0$ and $mr$ it is straightforward to find the corresponding $\lambda_0$ numerically. In the following table, we list a set of values of $\lambda_0$ for different values of $mr$.
\begin{center}
\begin{tabular}{|c|c|c|c|c|c|c|c|c|}
  \hline
  $mr$ & 3 & 4 & 5 & 6 & 7 & 8 & 9 & 10 \\
  \hline
  $\lambda_0$ & 6.78 & 5.60 & 4.86 & 4.36 & 4.01 & 3.75 & 3.55 & 3.39 \\
  \hline
\end{tabular}
\end{center}

\section{Non-unitary case : scaling Lee-Yang model}
\label{sec:nonunitary}
In 1+1 dimensions, non-unitary field theories make perfect physical sense and describe many statistical mechanics system. The computation of entanglement entropy for non-unitary IQFT's such as the scaling Lee-Yang model is more complicated than the unitary ones. First of all, the partition function on the Riemann surface is no longer given by the two-point function of branch-point twist fields. One needs to introduce a new kind of twist field which is defined by the operator product of the branch-point twist field $\mathcal{T}$ and the scalar field with smallest scaling dimension denoted by $\phi$. The partition function on the Riemann surface is then given by the ratio of these two kinds of two-point functions. Secondly, non-unitarity also affects spectral expansion since many fields become non-Hermitian. The computation of entanglement entropy in non-unitary IQFT's has been investigated in \cite{Bianchini:2015uea}. In this section, we consider non-unitary IQFT's with integrable topological defect using scaling Lee-Yang model as an example.

\subsection{Defect scaling Lee-Yang model}
The scaling Lee-Yang model can be obtained as a relevant perturbation of the CFT minimal model $\mathcal{M}_{(2,5)}$ with central charge $c=-22/5$. This model has only one type of neutral particle $\phi$ with mass $m$ and the bulk scattering is characterized by the following $S$-matrix
\begin{align}
S(\theta)=\frac{\sinh\theta+i\sin\frac{\pi}{3}}{\sinh\theta-i\sin\frac{\pi}{3}}
\end{align}
In this model, one can introduce the topological defect which preserves integrability with the transmission amplitude \cite{Bajnok:2007jg}
\begin{align}
\label{eq:TsLY}
T(\theta)=[b+1][b-1],\qquad [x]=i\frac{\sinh\left(\frac{\theta}{2}+i\frac{\pi\,x}{12}\right)}{\sinh\left(\frac{\theta}{2}+i\frac{\pi\,x}{12}-i\frac{\pi}{2}\right)}
\end{align}
The defect model and the corresponding form factor bootstrap program has been studied extensively, see for example \cite{Bajnok:2007jg,Bajnok:2009hp,Bajnok:2013waa}.

\subsection{Replica trick and entanglement entropy}
As in the unitary case, one can apply the replica trick to compute the entanglement entropy. However, the computation is more subtle due to non-unitarity. For non-unitary CFTs like Lee-Yang model, the physical vacuum state (\emph{i.e.} lowest energy state) does not coincide with conformal vacuum. Due to state-operator correspondence, we can associate the physical vacuum with a local field $\phi$ with scaling dimension $\Delta$. As expected, this field will enter the computation of EE. It is shown by careful analysis in \cite{Bianchini:2014uta,Bianchini:2015uea} that the partition function on the Riemann surface for non-unitary theories is now given by
\begin{align}
\label{eq:partitionLY}
\tr_{\mathcal{H}_A}\rho_A^n=\mathcal{Z}_n\,\varepsilon^{4(\Delta_{\mathcal{T}_\phi}-n\Delta)}
\frac{\langle\mathcal{T}_\phi(x_1)\tilde{\mathcal{T}}_\phi(x_2)\rangle}{\langle\phi(x_1)\phi(x_2)\rangle^n}
\end{align}
where we denote the scaling dimension of the branch-point twist field $\mathcal{T}$ as $\Delta_{\mathcal{T}}$. The normalization $\langle\phi(x_1)\phi(x_2)\rangle^n$ in the denominator can be seen as the norm of the vacuum state in CFT. The new twist field $\mathcal{T}_\phi$ is defined as
\begin{align}
\mathcal{T}_\phi(y)=n^{2\Delta-1}\,\lim_{x\to y}|x-y|^{2\Delta(1-\frac{1}{n})}\sum_{j=1}^n\mathcal{T}(y)\phi_j(x).
\end{align}
We denote the scaling dimension of this new branch-point twist field as $\Delta_{\mathcal{T}_\phi}$ which reads
\begin{align}
\label{eq:tLY}
\Delta_{\mathcal{T}_\phi}=\Delta_{\mathcal{T}}+\frac{\Delta}{n}=\frac{c}{24}\left(n-\frac{1}{n}\right)+\frac{\Delta}{n}
\end{align}
From (\ref{eq:partitionLY}) we see that in order to compute the entanglement entropy, we need to evaluate two types of two-point functions.

\subsection{Form factors and defect matrix elements}
In this subsection, we give the ingredient that are necessary for the the computation of the entanglement entropy. We need form factors of the twist field $\mathcal{T}_\phi$ and the fundamental field $\phi$ up to two particles. The form factor of field $\phi$ has been derived by Zamolodchikov \cite{Zamolodchikov:1990bk}.\\

\noindent\textbf{Form factors of fundamental field $\phi$}\\
The one-particle form factor is a constant due to relativistic invariance and is given by
\begin{align}
F^\phi_1=\frac{i\sqrt{2}}{3^{1/4}f(\frac{2\pi i}{3},1)}F_0^\phi
\end{align}
where
\begin{align}
F^\phi_0=\langle\phi\rangle=\frac{5i m^{-\frac{2}{5}}}{24h\sqrt{3}},\qquad h=0.09704845636
\end{align}
and the two particle form factor can be written as
\begin{align}
F_2^\phi(\theta)=\frac{\pi m^2}{8}\,\frac{F_{\text{min}}(\theta,1)}{f(i\pi,1)}
\end{align}
Here the two-point minimal form factor $F_{\text{min}}(\theta,n)$ is given by
\begin{align}
F_{\text{min}}(\theta,n)=a(\theta,n)f(\theta,n),
\end{align}
where $a(\theta,n)$ encodes the pole for bound states
\begin{align}
a(\theta,n)=\frac{\cosh\frac{\theta}{n}-1}{\cosh\frac{\theta}{n}-\cosh\frac{2\pi}{3n}}
\end{align}
and $f(\theta,n)$ is given by
\begin{align}
f(\theta,n)=\exp\left(2\int_0^\infty\frac{\sinh\frac{t}{3}\sinh\frac{t}{6}}{t\,\sinh(nt)\cosh\frac{t}{2}}\cosh t\left(n+\frac{i\theta}{\pi}\right)dt\right)
\end{align}
\\

\noindent\textbf{Form factors of twist field $\mathcal{T}_\phi$ and $\tilde{\mathcal{T}}_\phi$}\\
The one-particle form factor for the branch-point twist field is given by
\begin{align}
F_1^{\mathcal{T}_\phi|1}=\frac{\langle\mathcal{T}_\phi\rangle\,\Gamma}{2n\sin\left(\frac{\pi}{3n}\right)f\left(\frac{2\pi i}{3},n \right)},\qquad
\Gamma=i\sqrt{2}\,3^{1/4}
\end{align}
The two-particle form factor is given by
\begin{align}
F_2^{\mathcal{T}_\phi|11}(\theta)=\frac{\langle\mathcal{T}_\phi\rangle\,\sin\left(\frac{\pi}{n}\right)}
{2n\,\sinh\left(\frac{i\pi-\theta}{2n}\right)\sinh\left(\frac{i\pi+\theta}{2n}\right)}\frac{F_{\text{min}}(\theta,n)}{F_{\text{min}}(i\pi,n)}
+\frac{(F_1^{\mathcal{T}_\phi|1})^2}{\langle\mathcal{T}_\phi\rangle}F_{\text{min}}(\theta,n)
\end{align}
Form factors with generic replica indices can be expressed through the fundamental one by
\begin{align}
F_k^{\mathcal{T}_\phi|\mu_1\cdots\mu_k}(\theta_1,\cdots,\theta_k)=F_k^{\mathcal{T}_\phi|1\cdots1}(\theta_1+2\pi i(\mu_1-1),\cdots,\theta_k+2\pi i(\mu_k-1))
\end{align}
From the inversion of copy numbers, we can related the form factors of $\mathcal{T}_\phi$ and $\tilde{\mathcal{T}}_\phi$
\begin{align}
F_k^{\mathcal{T}_\phi|\mu_1\cdots\mu_k}(\theta_1,\cdots,\theta_k)=F_k^{\tilde{\mathcal{T}}_\phi|(n-\mu_1)\cdots(n-\mu_k)}(\theta_1,\cdots,\theta_k).
\end{align}
Non-unitarity of the theory modifies complex conjugation of the form factors by an extra phase factor
\begin{align}
\left[\langle\mathcal{T}_\phi\rangle^{-1}F_k^{\mathcal{T}_\phi|\mu_1\cdots\mu_k}(\theta_1,\cdots,\theta_k)\right]^*
=\textcolor{red}{(-1)^k}\langle\tilde{\mathcal{T}}_\phi\rangle^{-1}F_k^{\tilde{\mathcal{T}}_\phi|\mu_1\cdots\mu_k}(\theta_k,\cdots,\theta_1)
\end{align}
Note that we have $\langle\mathcal{T}_\phi\rangle=\langle\tilde{\mathcal{T}}_\phi\rangle$.
\\

\noindent\textbf{Matrix element of defect $\mathbb{D}$}\\
In the defect theory, we also need the matrix elements of the defect operator, which in our case is given by the transmission matrix (\ref{eq:tLY}).
In the computation of entanglement entropy, we need $\hat{T}(\theta)=T_-(\frac{i\pi}{2}-\theta)$. The defect depends on the parameter $b$. Here we consider a special case $b=3+i\,\alpha$ ($\alpha\in\mathbb{R}$) where the transmission amplitude $\hat{T}(\theta)$ is real and given by\footnote{This is also the regime where defect TBA is reliable, see \cite{Bajnok:2013waa}.}
\begin{align}
\label{eq:TsLYhat}
\hat{T}(\theta)=\frac{\cosh\left(\theta+\frac{\pi\alpha}{6}\right)+\frac{\sqrt{3}}{2}}{\cosh\left(\theta+\frac{\pi\alpha}{6}\right)-\frac{\sqrt{3}}{2}}
\end{align}
Our motivation for such a choice is that for generic parameter $b$, the correction to the entanglement entropy will be complex which has no obvious physical meaning. The one- and two-particle defect matrix element are given by the same formulas as in (\ref{eq:D22}). The main difference is that now both $\rD_{11}$ and $\rD_{22}$ will contribute to the entanglement entropy.

\subsection{Spectral expansion and entanglement entropy}
\begin{figure}[h!]
\begin{center}
\includegraphics[scale=0.45]{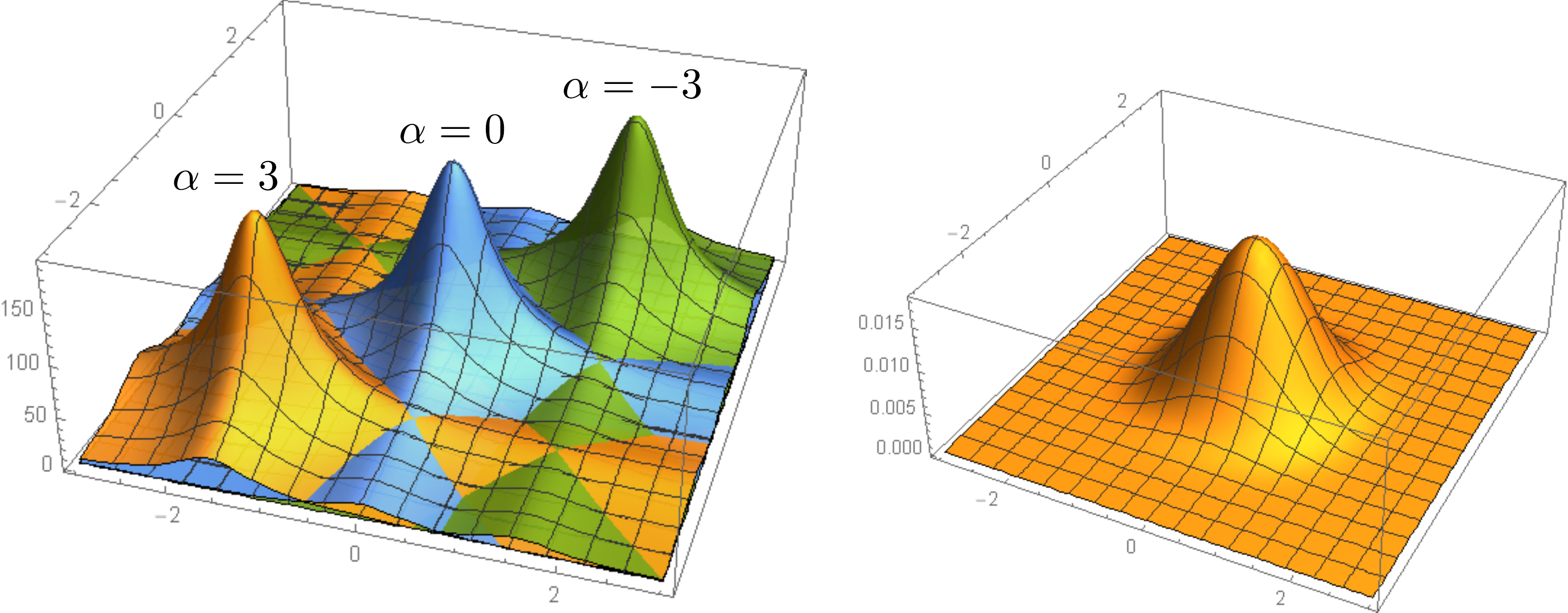}
\caption{Plot of $f_{\text{exp}}(\theta_1,\theta_2)$ and $f_\rD(\theta_1,\theta_2,\alpha)$. Here we take $mr=1$ and $\alpha=0,\pm3$.}
\label{fig:TT2}
\end{center}
\end{figure}

In this subsection, we compute the entanglement entropy using the form factor bootstrap method. Slightly generalizing the expression in \cite{Bianchini:2015uea}, we can write the entanglement entropy in terms of two-point functions as
\begin{align}
\label{eq:EEsLY}
S(r)=-\lim_{n\to 1}\frac{d}{dn}\left[\mathcal{Z}_n\,\varepsilon^{\frac{c_{\text{eff}}}{6}(n-\frac{1}{n})}
\frac{\langle\mathcal{T}_\phi\rangle^2}{\langle\phi\rangle^{2n}}\frac{A_\rD(r,n)}{B_\rD(r)^n}\right]
\end{align}
where
\begin{align}
A_\rD(r,n)=\frac{\langle\mathcal{T}_\phi(t_1)\mathbb{D}\tilde{\mathcal{T}}_\phi(t_2)\rangle}{\langle\mathcal{T}_\phi\rangle^2},\qquad
B_\rD(r)=\frac{\langle\phi(r)\mathbb{D}\phi(t_1)\rangle}{\langle\phi(t_2)\rangle^2}
\end{align}
and $r=|t_1-t_2|$. The entanglement entropy in the UV limit is not modified by the topological defect and is given by \cite{Bianchini:2014uta}
\begin{align}
S_A=\frac{c_{\text{eff}}}{3}\log(r/\varepsilon)
\end{align}
where for scaling Lee-Yang model $c_{\text{eff}}=c-24\Delta=4/5$. In the IR limit, the entanglement entropy (\ref{eq:EEsLY}) can be written as
\begin{align}
S(r)=-\frac{c_{\text{eff}}}{3}\log(\varepsilon m)+U-\lim_{n\to 1}\frac{d}{dn}\frac{A_\rD(r,n)}{B_\rD(r)^n}
\end{align}
where $U$ is defined in \cite{Bianchini:2015uea}. We concentrate on the leading correction which is given by
\begin{align}
\label{eq:ABexpand}
\lim_{n\to 1}\frac{d}{dn}\frac{A_\rD(r,n)}{B_\rD(r)^n}=\frac{A'_\rD(r,1)}{B_\rD(r)}-\log B_\rD(r)
\end{align}
where $A_\rD'(r,n)=d A_\rD(r,n)/dn$. Both $A_\rD(r,n)$ and $B_\rD(r)$ can be expanded in terms of form factors. We denote $A_{\rD,k}(r,n)$ and $B_{\rD,k}(r)$ the contribution of the $k$-particle state and
\begin{align}
A_{\rD}(r,n)=&\,A_{\rD,1}(r,n)+A_{\rD,2}(r,n)+\cdots\\\nonumber
B_{\rD}(r)=&\,B_{\rD,1}(r)+B_{\rD,2}(r)+\cdots
\end{align}
Keeping the r.h.s of (\ref{eq:ABexpand}) up to two particle contributions, we obtain
\begin{align}
\lim_{n\to 1}\frac{d}{dn}\frac{A_\rD(r,n)}{B_\rD(r)^n}=&\,A'_{\rD,1}(r,1)+A'_{\rD,2}(r,1)-B_{\rD,1}(r)-B_{\rD,2}(r)\\\nonumber
&\,+\frac{1}{2}B_{\rD,1}^2-A'_{\rD,1}(r,1)B_{\rD,1}(r)+\cdots
\end{align}
Now we compute the above quantities
\begin{align}
A_{\rD,1}(r,n)=-n\int_{-\infty}^{\infty}\frac{d\theta}{2\pi}\left|\frac{F_1^{\mathcal{T}_\phi|1}}{\langle\mathcal{T}_\phi\rangle}\right|^2
\hat{T}(\theta)
e^{-mr\cosh\theta}
\end{align}
Note that the factor $F^{\mathcal{T}_\phi|1}_1$ contains $n$ in a non-trivial way, so that $A'_{\rD,1}(r,1)$ needs to be computed with some care and we obtain
\begin{align}
\label{eq:AD1}
A'_{\rD,1}(r,1)=&\,-\frac{13}{108 f(\frac{2\pi i}{3},1)^2}\int_{-\infty}^\infty\hat{T}(\theta)e^{-mr\cosh\theta} d\theta\\\nonumber
=&\,-\frac{13}{108 f(\frac{2\pi i}{3},1)^2}\int_{-\infty}^\infty\left( \frac{\cosh\left(\theta+\frac{\pi\alpha}{6}\right)+\frac{\sqrt{3}}{2}}{\cosh\left(\theta+\frac{\pi\alpha}{6}\right)-\frac{\sqrt{3}}{2}} \right)e^{-mr\cosh\theta} d\theta
\end{align}
The derivative of the two particle contribution takes a simpler form due to the general result (\ref{eq:anacon}). The two-particle contribution reads
\begin{align}
A_{\rD,2}(r,n)=\frac{1}{2}\sum_{i,j=1}^n\int_{-\infty}^{\infty}\frac{d\theta_1d\theta_2}{(2\pi)^2}\left| \frac{F_2^{\mathcal{T}_\phi|ij}(\theta_1,\theta_2)}{\langle\mathcal{T}_\phi\rangle} \right|^2\hat{T}(\theta_1)\hat{T}(\theta_2) e^{-mr\cosh\theta_1-mr\cosh\theta_2}.
\end{align}
Using (\ref{eq:anacon}), the derivative of $A_{\rD,2}(r,n)$ at $n=1$ is given by
\begin{align}
\label{eq:AD2}
A'_{\rD,2}(r,1)=&\,\frac{1}{16}\int_{-\infty}^{\infty}d\theta\, \hat{T}(\theta)^2 e^{-2mr\cosh\theta}\\\nonumber
=&\,\frac{1}{16}\int_{-\infty}^{\infty}d\theta
\left( \frac{\cosh\left(\theta+\frac{\pi\alpha}{6}\right)+\frac{\sqrt{3}}{2}}{\cosh\left(\theta+\frac{\pi\alpha}{6}\right)-\frac{\sqrt{3}}{2}} \right)^2
 e^{-2mr\cosh\theta}
\end{align}
The one-particle contribution for $B_{\rD,1}(r)$ is
\begin{align}
\label{eq:BD1}
B_{\rD,1}(r)=&\,-\int_{-\infty}^\infty\frac{d\theta}{2\pi}\left|\frac{F_1^\phi}{\langle\phi\rangle}\right|^2\hat{T}(\theta) e^{-mr\cosh\theta}\\\nonumber
=&\,-\left|\frac{F_1^\phi}{\langle\phi\rangle}\right|^2\int_{-\infty}^{\infty}\frac{d\theta}{2\pi}\left(
\frac{\cosh\left(\theta+\frac{\pi\alpha}{6}\right)+\frac{\sqrt{3}}{2}}{\cosh\left(\theta+\frac{\pi\alpha}{6}\right)-\frac{\sqrt{3}}{2}}\right)\, e^{-mr\cosh\theta}
\end{align}
The two-particle contribution reads
\begin{align}
\label{eq:BD2}
B_{\rD,2}(r)=\frac{1}{2}\int_{-\infty}^\infty\frac{d\theta_1 d\theta_2}{(2\pi)^2}\left|\frac{F_2^{\phi}(\theta_1-\theta_2)}{\langle\phi\rangle}\right|^2
\hat{T}(\theta_1)\hat{T}(\theta_2)\,e^{-mr\cosh\theta_1-mr\cosh\theta_2}.
\end{align}
Putting all the terms together, we write down the final result in the IR limit
\begin{align}
\label{eq:Sdefect_non-unitary}
S_A^{\text{defect}}=&\,-\frac{2}{15}\log(\varepsilon m)+U-\left(\frac{13}{108f(\frac{2\pi i}{3},1)}
+\left|\frac{F_1^{\phi}}{\langle\phi\rangle}\right|^2\right)\int_{-\infty}^{\infty}\hat{T}(\theta)e^{-mr\cosh\theta}d\theta\\\nonumber
&\,+\left|\frac{F_1^{\phi}}{\langle\phi\rangle}\right|^2\left(\frac{13}{108f(\frac{2\pi i}{3},1)}
+\left|\frac{F_1^{\phi}}{\langle\phi\rangle}\right|^2  \right)\left( \int_{-\infty}^{\infty}\hat{T}(\theta)e^{-mr\cosh\theta}d\theta \right)^2\\\nonumber
&\,-\frac{1}{2}\int_{-\infty}^{\infty}\left|\frac{F_2^\phi(\theta_1-\theta_2)}{\langle\phi\rangle}\right|^2\hat{T}(\theta_1)\hat{T}(\theta_2)
e^{-mr(\cosh\theta_1+\cosh\theta_2)}d\theta_1d\theta_2\\\nonumber
&\,+\frac{1}{16}\int_{-\infty}^{\infty}\hat{T}(\theta)^2 e^{-2mr\cosh\theta}d\theta+\cdots
\end{align}
The first three quantities (\ref{eq:AD1}),(\ref{eq:AD2}) and (\ref{eq:BD1}) take the form that is similar to (\ref{eq:leading}) and the same analysis as the unitary case applies. For the quantity $B_{\rD,2}$ we need to integrate over two variables. Although technically more involved, physically it is the same as the unitary case, namely the exponential factors $f_{\text{exp}}(\theta_1,\theta_2)=e^{-mr\cosh\theta_1-mr\cosh\theta_2}$ and the defect matrix element $f_\rD(\theta_1,\theta_2,\alpha)=\hat{T}(\theta_1)\hat{T}(\theta_2)$ have certain peaks in a finite region and quickly damp away outside, as is shown in figure\,\ref{fig:TT2}. The position of the peak is controlled by the parameter $\alpha$.
If the peaks of $f_\text{exp}$ and $f_\rD$ overlap, the defect has finite contribution to entanglement entropy, otherwise the contribution is negligible and we recover the bulk result.

Similar to the unitary case, we give numeric estimation of the peak overlapping. From our final expression (\ref{eq:Sdefect_non-unitary}), there are two types of overlaps to consider, namely the overlap of $f_{\text{exp},1}=e^{-mr\cosh\theta}$, $f_{\rD,1}=\hat{T}(\theta)-1$ (from the first two lines) and of $f_{\text{exp},2}=e^{-2mr\cosh\theta}$, $f_{\rD,2}=\hat{T}(\theta)^2-1$ (from the last two lines) where $\hat{T}(\theta)$ is given by (\ref{eq:TsLYhat}). As in the unitary case, we set $\alpha=\lambda\,mr$. For the two cases, we have $\text{I}_{\text{max},1}=e^{-mr}$ and $\rI_{\text{max},2}=e^{-2mr}$. We set $\eta_0^*=1.0\times 10^{-5}$ as before for both cases, the corresponding $\lambda_0$ are given in the following tables. For the first case, we have
\begin{center}
\begin{tabular}{|c|c|c|c|c|c|c|c|c|}
  \hline
  $mr$ & 3 & 4 & 5 & 6 & 7 & 8 & 9 & 10 \\
  \hline
  $\lambda_0$ & 8.59 & 7.03 & 6.07 & 5.41 & 4.94 & 4.58 & 4.30 & 4.07 \\
  \hline
\end{tabular}
\end{center}
For the second case, we have
\begin{center}
\begin{tabular}{|c|c|c|c|c|c|c|c|c|}
  \hline
  $mr$ & 3 & 4 & 5 & 6 & 7 & 8 & 9 & 10 \\
  \hline
  $\lambda_0$ & 11.27 & 9.50  & 8.42 & 7.68 & 7.15 & 6.75 & 6.44 & 6.19 \\
  \hline
\end{tabular}
\end{center}

\section{Conclusions}
\label{sec:conclude}
In this paper, we considered the effect of integrable purely transmissive line defect on the bipartite entanglement entropy in 1+1 dimensional integrable field theories. We show that the topological defect does not affect the universal scaling of entanglement entropy in the UV limit. In the IR limit, the line defects have finite contribution to the leading corrections of entanglement entropy within certain range of the parameters that characterize the defect and have no contribution outside this finite range. The result holds for both unitary and non-unitary theories.\par

As a future direction, it is very interesting to investigate the finite volume/temperature entanglement entropy for integrable field theories, first for the bulk case and then for the boundary/defect case. Although obtaining the result for generic volume/temperature might be quite challenging, a systematic low temperature/large volume expansion for the bulk case should be within reach following similar ideas proposed in \cite{Pozsgay:2007kn,Pozsgay:2007gx,Pozsgay:2010cr}. The generalization to the defect case is also known \cite{Bajnok:2013eaa}.

In the bulk case, the leading exponential correction of the entanglement entropy is universal even for \emph{non-integrable} massive field theories \cite{Doyon:2008vu}. In this paper, we have shown that the leading exponential correction is also universal for the topological defect case. It would be interesting to see if we could generalize this result to generic non-integrable massive field theories\footnote{We thank Benjamin Doyon for suggesting this point.}.

\section*{Acknowledgements}
\label{sec:conclude}
I would like to thank Juan Jottar for initial collaborations on the project. I'm also indebted to Zoltan Bajnok and Laszlo Hollo for many helpful discussions and correspondences about integrable defects. I also thank Zoltan Bajnok, Song He and Benjamin Doyon for very helpful comments on the manuscript. This work is partially supported by the Swiss National Science Foundation through the NCCR SwissMap.

\providecommand{\href}[2]{#2}\begingroup\raggedright\endgroup


\end{document}